\documentstyle[preprint,aps]{revtex}
\begin{document}

\preprint{\vbox{\hbox{UCD-97-26} }}
\draft
\title {$V-A$ Constraint on a Product of $R$-parity Violating Couplings}
\author{ Kingman Cheung and Ren-Jie Zhang}
\address{
Department of Physics, University of California at Davis, Davis, 
CA 95616}
\date{December, 1997}

\maketitle

\begin{abstract} 
We study in the framework of $R$-parity violating supersymmetric theories
the effect of $R$-parity violation due to the operator $L_i L_j\overline{E_k}$
on the $(V-A)$ structure of the muon
decay.  The precisely measured muon decay parameters can constrain 
a product of $R$-parity violating couplings:
$|\lambda_{232} \lambda_{131}|< 0.022$ at the 90\% CL,
which is complementary to the previous limits obtained by the $e$-$\mu$
universality in $\tau$ decay.
\end{abstract}

\section{Introduction} 

$R$-parity violation is introduced into the minimal supersymmetric 
standard model  through additional terms in the superpotential:
\begin{equation}
{\cal W}_{\overlay{/}{\scriptstyle R}} = 
\lambda_{ijk} L_i L_j \overline{E_k} + \lambda'_{ijk}
 L_i Q_j \overline{D_k} + \lambda^{''}_{ijk} \overline{U_i} \,
\overline{D_j} \, \overline{D_k} +
\mu_i L_i H_u \;,
\end{equation}
where $L,{\overline E},Q,{\overline U},{\overline D}, H_u$ 
are the superfields,  $i,j,k$ are family indices, and 
$\lambda,\; \lambda'$ and $\lambda^{''}$ are the $R$-parity violating
(RPV) couplings.   The last term $\mu_i L_i H_u$ 
can be rotated away by a redefinition of the lepton field.
(We neglect the effects of a possible  soft-breaking bilinear term.)
The operators $LL\overline{E}$ and $LQ\overline{D}$ violate lepton number
while $\overline{U}\, \overline{D}\, \overline{D}$ 
violates baryon number.  A prominent  constraint
coming from proton decay requires either $\lambda'$ or $\lambda^{''}$
to be zero.  Moreover, these RPV couplings would violate a number of 
existing data.  The present limits on  these RPV couplings are 
listed in recent reviews \cite{dreiner},
where the limits are obtained by assuming only one nonzero coupling 
at a time.  Constraints on products of RPV couplings have also been
calculated for proton stability, lepton-family-number violating processes,
and flavor-changing neutral current (FCNC) processes \cite{bh}.

In this note, we are primarily interested in the $\lambda L L \overline{E}$
term.  We point out that the precise measurement on the 
$(V-A)$ structure in $\mu$ decay puts an additional constraint on 
a product of $\lambda$s, namely $\lambda_{131} \lambda_{232}$.
As will be shown later, using the $e$-$\mu$ universality in $\tau$ decay 
to put limits on $\lambda_{i3k}$ might run into the danger
if several couplings coexist;
in particular, when $|\lambda_{13k}|$ and $|\lambda_{23k}|$ 
are approximately equal their contributions to $R_\tau$ cancel.
Thus, in this case  the $e$-$\mu$ universality cannot 
effectively constrain the $\lambda$s; however, 
the constraint on the product of the two $\lambda$s, 
$\lambda_{131} \lambda_{232}$,
from the $(V-A)$ structure in $\mu$ decay remains useful.
Note that all previous constraints on 
products of RPV couplings come from FCNC processes or lepton-family-number
violating processes; here the $(V-A)$ structure in $\mu$ decay 
does not involve any of these.

The $(V-A)$ structure has been tested
in a number of processes, e.g., $\pi$ decay, $\tau$ decay, $\mu$ decay,
of which the $\mu$ decay was measured to a very high precision.  In the
following, we will use the $\mu$ decay parameters to constrain the product
$|\lambda_{131} \lambda_{232}|$.  We find that our new  limit is complementary
to the previous limit obtained using the $e$-$\mu$
universality in the $\tau$ decay: $\tau^- \to e^- \bar \nu_e \nu_\tau$,
$\tau^- \to \mu^- \bar \nu_\mu \nu_\tau$.  We shall first describe the 
general structure of the $\mu$ decay.  Next, we shall derive the effect of
the RPV couplings and obtain the upper limit on the $\lambda$s.
Finally, we comment on the constraint coming from the high energy process
$e^+ e^- \to \mu^+ \mu^-$ and conclude.

\section{Muon Decay Parameters}

The muon decay and the inverse muon decay at low energy can be conveniently 
parameterized in terms of amplitudes $g_{\epsilon\mu}^\gamma$ and the 
Fermi Constant $G_F$, using the matrix element \cite{fetscher}
\begin{equation}
\frac{4G_F}{\sqrt{2}} \sum_{\stackrel{\scriptstyle \gamma=V,S,T}
{\epsilon,\mu=L,R}} g_{\epsilon\mu}^\gamma \;
\langle \overline{e_\epsilon} | \Gamma^\gamma | (\nu_e)_n \rangle \;
\langle \overline{(\nu_\mu)_m}| \Gamma_\gamma | \mu_\mu \rangle  \;,
\end{equation}
where $\gamma=V,S,T$ denotes a vector, scalar, or tensor interaction,
$\epsilon,\mu$ denote the chirality of the electron and muon, respectively,
and the chiralities $n$ and $m$ of $\nu_e$ and $\overline{\nu_\mu}$ are
determined by $\gamma,\epsilon,\mu$.  
In the standard model, the $(V-A)$ requires $g_{LL}^V=1$ and others equal 
zero.  The rate, energy and angular distributions, and polarization can
be affected by these $g_{\epsilon,\mu}^\gamma$.  In the rest frame of 
the muon, the energy and angular distribution is given by the Michel
spectrum:
\begin{equation}
\frac{d^2 \Gamma}{dx d\cos\theta} \sim
\left\{ 3(1-x) + \frac{2\rho}{3} (4x-3) \mp \xi \cos\theta \left[
1-x +\frac{2\delta}{x}(4x-3) \right ] \right \} x^2 \;,
\end{equation}
where $\rho, \xi,\delta$ are functions of $g_{\epsilon,\mu}^\gamma$.  
The measurements of $\rho, \xi,\delta$ can constrain various combinations of
$g_{\epsilon,\mu}^\gamma$.   In order to determine the 
amplitudes $g_{\epsilon,\mu}^\gamma$ uniquely, Fetscher et al. introduced
four probabilities $Q_{\epsilon\mu} (\epsilon,\mu=L,R)$ for the decay of
a $\mu$-handed muon into a $\epsilon$-handed electron \cite{fetscher}:
\begin{equation}
Q_{\epsilon\mu} = \frac{1}{4} \left |g_{\epsilon,\mu}^S \right|^2 +
 \left |g_{\epsilon,\mu}^V \right|^2 + 3(1- \delta_{\epsilon,\mu} )
\left |g_{\epsilon,\mu}^T \right|^2 \;.
\end{equation}
The $Q_{LL}$ is constrained to be very close to unity, while others very close
to zero.  
The current limits on $g_{\epsilon,\mu}^\gamma$ 
are summarized in the Particle Data Book \cite{PDG}.  
The ones that are relevant to our analysis are 
\begin{equation}
\label{expt}
|g_{RR}^S | < 0.066\;, \qquad |g_{LL}^V| > 0.96  
\end{equation}
at the 90\% CL.

\section{Effect of $R$ parity violation}

With the term $\frac{1}{2} \lambda_{ijk} L_i L_j \overline{E_k}$ in the 
superpotential  the Lagrangian is given by
\begin{equation}
{\cal L} = \lambda_{ijk} \left\{ 
\tilde{e}^*_{kR} \overline{(\ell_{jL})^c}\nu_{iL}  
+ \overline{e_{kR}} \nu_{iL} \tilde{\ell}_{jL} 
- \overline{e_{kR}} \ell_{iL} \tilde{\nu}_{jL} 
\right \} + h.c.
\label{lag}
\end{equation}
There are two possible diagrams contributing to the muon decay.  The first
one is via an exchange of $\tilde\tau_L$  and the amplitude is given by
\begin{equation}
{\cal L}_1 = - \,\frac{\lambda_{131} \lambda^*_{232}}{m^2_{\tilde{\tau}_L}} \;
\left( \overline{e_R} \nu_{eL} \right ) \; 
\left( \overline{\nu_{\mu L}} \mu_R \right ) \; .
\end{equation}
This amplitude contributes to $g_{RR}^S$ as follows
\begin{equation}
\label{8}
\delta \left( g_{RR}^S \right) = -\, \frac{\sqrt{2}}{4G_F} \;
\frac{\lambda_{131} \lambda^*_{232} }{m^2_{\tilde{\tau}_L} } \;.
\end{equation}
Note that this contribution has a different helicity structure as the SM 
$(V-A)$ amplitude and, therefore, the experimental limit on the $(V-A)$
structure can effectively constrain the product 
$|\lambda_{131} \lambda_{232}|$.  
Using Eqs. (\ref{expt}) and (\ref{8}) we obtain at the 90\% CL, for 
$m_{\tilde{\tau}_L}=100$ GeV,
\begin{equation}
\label{our}
|\lambda_{131} \lambda_{232} | < 0.022 \;.
\end{equation}

The second one is via an exchange of $\tilde{e}_R, \tilde{\mu}_R$, 
or $\tilde\tau_R$.  The amplitude
is given by
\begin{equation}
{\cal L}_2 = -\,  \sum_{k=1}^3
\frac{|\lambda_{12k} |^2}{2 m^2_{\tilde{\ell}_{kR}}} \;
\left( \overline{e_L} \gamma^\mu \nu_{eL} \right ) \; 
\left( \overline{\nu_{\mu L}} \gamma_\mu \mu_L \right)  \; .
\end{equation}
This amplitude contributes to $g_{LL}^V$:
\begin{equation}
\delta \left( g_{LL}^V \right )= -\,\frac{\sqrt{2}}{4G_F} \;
\sum_{k=1}^3 \frac{|\lambda_{12k}|^2}{2 m_{\tilde{\ell}_{kR}}^2} \;.
\end{equation}
This ${\cal L}_2$ has the same helicity structure as the SM $(V-A)$ amplitude
and, therefore, the $(V-A)$ structure cannot constrain $|\lambda_{12k}|^2$,
but the total rate should be able to do so (similar to the analysis in
\cite{hagi}.)
However, it was shown \cite{bgh} that the $e$-$\mu$-$\tau$ universality 
is also able to constrain $|\lambda_{12k}|^2$ to a very small value.

Recall that the previous constraints on $\lambda_{13k}$ and $\lambda_{23k}$
came from the $e$-$\mu$ universality in $\tau$ decay:
\begin{equation}
\label{rtau}
R_\tau \equiv \frac{\Gamma(\tau \to e \nu \bar \nu)}
                   {\Gamma(\tau \to \mu \nu \bar \nu)} =
 R^{\rm SM}_\tau \left[ 1+ \frac{1}{2\sqrt{2} G_F} \;
\sum_{k=1}^3 \left( \frac{|\lambda_{13k}|^2}{m^2_{\tilde{\ell}_{kR}}} -
       \frac{|\lambda_{23k}|^2}{m^2_{\tilde{\ell}_{kR}}}  \right ) \right ]\;.
\end{equation}
The constraint on each $\lambda_{i3k}$ was obtained from the experimental
value of $R_\tau=1.0006\pm0.0103$ \cite{aleph} 
assuming only one $\lambda$ nonzero
at a time.  The limit  was $|\lambda_{i3k}|<0.076$ at 90\% CL \footnote{
In the reviews \cite{dreiner} only the $1\sigma$ results are given.
In order for a direct comparison with our result we convert their limits
to $1.65\sigma$ level, i.e., 90\% CL.  In the PDB only 90\% CL upper
bounds on $g_{\epsilon\mu}^\gamma$ are listed.}
for $m_{\tilde{\ell}_{kR}}=
100$ GeV and $i=1,2$ and $k=1,2,3$.  The danger of this limit can be seen
from Eq. (\ref{rtau}).  When $|\lambda_{13k}| \approx 
|\lambda_{23k}|$ their contributions to $R_\tau$ cancel and, therefore,  
the limits on $|\lambda_{i3k}|$ are no longer valid.  Physically, 
if the partial widths of the tau into muon and electron increased with
the same amount, the $e$-$\mu$ universality in the tau decay could not 
constrain the $\lambda$s.

The importance of our result in Eq. (\ref{our})  can be appreciated
immediately.  Even in the scenario where $|\lambda_{131}|\approx 
|\lambda_{232}|$ 
(the $e$-$\mu$ universality in tau decay is not useful anymore)
our result in Eq. (\ref{our}) can constrain them effectively to be 
$|\lambda_{131}|=|\lambda_{232}| < 0.15$ at 90\% CL.   
Of course, when $\lambda_{131}$ is very different from
$\lambda_{232}$  the limit from $e$-$\mu$ universality is more 
restrictive.

Actually, we can combine our result in Eq. (\ref{our}), the limit from
$e$-$\mu$ universality in $\tau$ decay, and the limit from $e$-$\mu$-$\tau$
universality in 
$\Gamma(\tau \to \mu \nu \bar \nu)/\Gamma(\mu \to e \nu \bar \nu)$
(which constrains $\sum [ |\lambda_{12k}|^2 - |\lambda_{23k}|^2]$),
as well as other constraints coming from the 
lepton-family-number violating processes and FCNC processes \cite{bh}, 
e.g., $\tau \to 3e$ constrains $|\lambda_{121} \lambda_{123}|, 
|\lambda_{131} \lambda_{133}|, |\lambda_{231} \lambda_{121}|$, 
$\mu \to 3e$ constrains $|\lambda_{121} \lambda_{122}|, |\lambda_{131} 
\lambda_{132}|, |\lambda_{231} \lambda_{131}|$, etc.
These constraints on products of two different $\lambda$s are 
complementary to the constraints obtained by the $e$-$\mu$-$\tau$
universality.  
In principle, we can perform a combined statistical analysis using all these
constraints to find a global set of constraints on all these RPV couplings
with correlations.

\section{Discussions}

There will be future experiments on measuring the muon decay parameters
with better sensitivity.  A planned experiment TRIUMF-E614 \cite{614}
is scheduled
to run and will have a sensitivity of $\rho,\delta, \xi$ down to $10^{-4}$.
Such sensitivity on $\rho,\delta,\xi$ will be able to pin  $|g_{RR}^S|$
down to $10^{-2}$, which would then give the limit on $|\lambda_{131}|=
|\lambda_{232}|$:
\begin{equation}
|\lambda_{131}| = |\lambda_{232}| \alt 0.06 \;.
\end{equation}

The product $|\lambda_{131}\lambda_{232}|$ can also be constrained
by high energy experiments at $e^+ e^-$ and $\mu^+ \mu^-$ colliders
\cite{kal,L3,feng}. 
The Lagrangian of Eq. (\ref{lag}) also contributes to the process 
$e^+e^-\rightarrow \mu^+\mu^-$ via $s$-channel exchanges of 
${\tilde\nu_{\tau L}}$ and ${\tilde\nu^*_{\tau L}}$ 
(depending on the coupling, there could also be $t$-channel diagrams).
The inclusion of scalar tau-sneutrino exchanges will affect both the cross
section and the forward-backward asymmetry in muon-pair production.
The change in cross section due to $s$-channel resonance production 
is given by
\begin{equation} 
\delta\sigma\ =\ \frac{|\lambda_{131}\lambda_{232}|^2}{32\pi}
\ \frac{s} { (s-m^2_{\tilde\nu_{\tau L}})^2 + \Gamma^2_{\tilde\nu_{\tau L}}
m^2_{\tilde\nu_{\tau L}} }\ .
\end{equation}
If the product of $\lambda$s is of an appreciable size and the mass of
the scalar tau-sneutrino is below the energy of the machine, the LEP2 and
the future NLC experiments should be able to see a prominent peak
by scanning over 
the center-of-mass energy (LEP2 has effectively done that due to
the initial state radiation); otherwise, the null result should be able to 
constrain the product of $\lambda$s.
If the mass of the scalar tau-sneutrino is above the center-of-mass energy
of the machine, only the effect from the tail of the scalar tau-sneutrino 
can be seen and, therefore, the limit on $\lambda$s is much weaker.
The L3 collaboration has recently published the 90\% CL upper limit on
$|\lambda_{131}|=|\lambda_{232}| \alt 0.04$ for $m_{\tilde{\nu}_{\tau L}}=
110 - 170$ GeV by measuring the cross section and the forward-backward
asymmetry in muon-pair production \cite{L3}.
The future experiment at the NLC can probe heavier scalar tau-sneutrino with
$|\lambda_{131} \lambda_{232}|$  down to $10^{-4}$ level, assuming 
$\Gamma_{\tilde\nu_{\tau L}}/m_{\tilde\nu_{\tau L}}\sim 1\%$
and an integrated luminosity of $\sim 50$ fb$^{-1}$.  

To conclude we have obtained a limit on
$|\lambda_{131} \lambda_{232}|< 0.022$ 
(or $|\lambda_{131}|=|\lambda_{232}|<0.15$) at the 90\% CL from the 
$(V-A)$ measurement
in the muon decay.  Although this limit is not as good as 
the previous limits $|\lambda_{13k}|,|\lambda_{23k}|<0.076$ for 
$m_{\tilde{\ell}_{kR}}=100$ GeV at 90\% CL obtained by 
the $e$-$\mu$ universality in $\tau$ decay,   our limit is, however, 
very useful for the case when $|\lambda_{131}| \approx |\lambda_{232}|$,
in which case the $e$-$\mu$ universality is satisfied no matter how large the
$\lambda$s are. It should be 
noted that the scenario of several coexisting $R$-parity violating couplings
is more complicated than the case previously examined in the
literature, and one should extract as much information as possible
from the existing experiments.

This research was supported in part by the U.S.~Department of Energy 
under Grant No. DE-FG03-91ER40674.

\end{document}